\documentstyle[aps,multicol,epsfig]{revtex}
\begin{document}
\draft
\title{Thermodynamic properties of a simple model\\
of like-charged attracting rods}
\author{J\"urgen F. Stilck}
\address{Instituto de F\'{\i}sica, Universidade Federal Fluminense\\
Av. Litor\^anea s/n, 24210-340, Niter\'oi, RJ, Brazil}
\author{Yan Levin and Jeferson J. Arenzon}
\address{ Instituto de F\'{\i}sica, Universidade Federal do Rio
Grande do Sul\\
Caixa Postal 15051, 91501-970, Porto Alegre, RS, Brazil}
\date{\today}
\maketitle
\begin{abstract}
We study the thermodynamic properties of a simple model for the
possible mechanism of attraction between like charged
rodlike polyions inside a polyelectrolyte solution. We consider two
polyions in parallel planes, with $Z$ charges each, in a solution 
containing multivalent counterions of valence $\alpha$. 
The model is solved exactly for $Z
\le 13$ for a general angle $\theta$ between the rods and supposing
that $n$ counterions are condensed on each polyion. The free energy
has two minima, one at $\theta=0$ (parallel rods) and another at
$\theta=\pi/2$ (perpendicular rods). In general, in situations where
an attractive force develops at small distances between the centers
of the polyions, the perpendicular configuration has the lowest free
energy at large distances,  while at small distances the parallel
configuration minimizes the free energy of the model. However, at low
temperatures, a reentrant behavior is observed, such that the
perpendicular configuration is the global minimum for both large and small
distances, while the parallel configuration minimizes the free energy
at intermediate distances. 
\end{abstract}


\begin{multicols}{2}
\section{Introduction}
\label{I}

It has been observed that like-charged macromolecules can
attract each other in solutions containing multivalent counterions.
This attraction manifests itself in {\it in vitro} 
formation of toroidal bundles of concentrated DNA~\cite{blo91,blo97}, 
similar to the
one found in  bacteriophage heads~\cite{kli67}, and the appearance of  
rodlike bundles of  f-actin and tobacco
mosaic virus in the presence of  multivalent counterions~\cite{tan96}. 
A number of models have been suggested to explain this curious
phenomenon.  The fundamental ingredient in all of these models
is the role played by the condensed counterions~\cite{pat80,oos68}.  
Thus, the attraction
has been attributed to the correlations between the condensed
counterions on the two polyions~\cite{ha97,rou96}.  The mathematical problem
of how these correlations can be taken into account is highly
non-trivial.  Two approaches have been proposed.  One relies
on  field theoretic methodology and uses what can be classified
as a high temperature expansion to account for the correlations
between the condensed counterions~\cite{ha97}, while the second is a zero
temperature approximation, in which the counterions are thought to
form correlated Wigner crystals on the surfaces of the two 
polyions~\cite{rou96,are99,sol99,kor99}.
Neither one of the approaches is exact, although the zero temperature
Wigner crystal approximation seems to be better at capturing
the true nature of correlations~\cite{lev99}.  

In a previous work we have introduced a simple model which has
allowed us to study {\it exactly} the force between two parallel charged rods 
with a layer of condensed counterions~\cite{are99}.  In this paper
we shall extend this work to allow for a relative inclination
between the two rods.  This problem is of particular interest
in kinetics of the bundle formation. Indeed, it has been
observed that the bundles of stiff polyions have a characteristic
size.  This is quite surprising since the correlation induced
attraction  should favor the
formation of infinitely thick bundles, after all the ionic crystals
can grow to macroscopic sizes.  A possible explanation for the
bundles not growing beyond some specific size can be found in the
kinematics of bundle formation.  The condensed counterions do not
fully neutralize the charge of a polyion.  For a rodlike polyion formed by
$Z$ $-q$ charges separated by distance $b$, placed in a solvent with
dielectric constant $D$, in the presence of counterions of charge
$\alpha q$, the Manning criterion~\cite{man69} states that the number
of condensed counterions on a polyion is $n=(1-1/\alpha\xi)Z/\alpha$
where $\xi=q^2/Dbk_BT$.  Thus, for the case of DNA with divalent
counterions $88\%$ of the DNA's charge is neutralized.  It is easy to
convince oneself that if the interaction between two rodlike
molecules is repulsive, there will be a greater probability that they
will be found  perpendicular to one another.  The correlation induced
attraction between the DNA molecules is short ranged, what means
that the electrostatic repulsion is dominant on large scales.  Thus,
the two polyions will in general repel one another.  It is only when
the two macromolecules come in a close contact that the thermal
fluctuations might be able to overcome the free energy barrier
between the perpendicular and the parallel configurations, allowing
the correlation induced attraction to take over and the polyions to
``bundle up''. It has been argued that the size of this free energy
barrier scales with the size of the bundle already
formed~\cite{ha99}.  Thus, there comes a point when the thermal
fluctuations will not be able to overcome the free energy barrier.
Motivated by this discussion we shall now proceed with the study of
interactions between rotating like-charged rods.

\section{Definition of the model and its solution}
\label{II}

\begin{figure}
\centerline{\epsfig{file=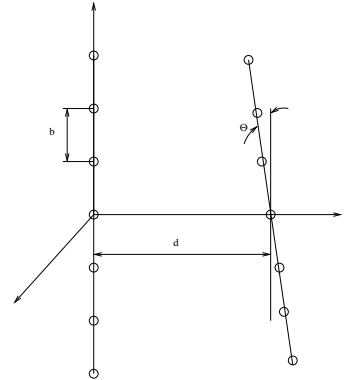,width=8cm,angle=270} }
\caption{Two rodlike polyions with $z=7$ charges each.}
\label{f1}
\end{figure}

We consider two rodlike polyions, each one with $Z$ charges $-q$ with
separation $b$ between neighbor charges. On each polyion,  $n$
$\alpha$-valent counterions, with charge $\alpha q$,
 are condensed. The sites where
the condensed counterions are located are described by occupation variables
$\sigma_{ij}$, such that $\sigma_{ij}=1$ if a counterion is
condensed on the $i$'th monomer ($i=1,\ldots,Z$) of the $j$'th polyion
($j=1,2$) and $\sigma_{ij}=0$ otherwise. When a counterion is
condensed on a monomer, we assume that the only effect is the
renormalization of the local charge, from $-q$ to
$-q+\alpha q$. The rods are located on
two parallel planes separated by a distance $d$ and the line
joining the centers of the rods is supposed to be perpendicular to
these planes. The angle between the directions of the rods is equal
to $\theta$, so that $\theta=0$ corresponds to the case of parallel
rods previously considered~\cite{are99}. 
The definitions above are illustrated in Fig.~\ref{f1}.
Notice that, for simplicity, we will consider only odd values of
$Z$, so that for nonzero values of $\theta$ the distance between the
central charges of each rod vanishes as $d \to 0$.
The polyions are placed in a uniform solvent whose dielectric
constant is equal to $D$. For a given configuration $\{\sigma \}$ of 
condensed counterions the Hamiltonian for the pair of polyions
may then be written as
\begin{equation}
{\cal H}=\frac{q^2}{2D}\sum_{i,i^{'}=1}^Z\sum_{j,j^{'}=1}^2
\frac{(1-\alpha \sigma_{ij})(1-\alpha \sigma_{i^{'}j^{'}})} 
{b\delta(i,j,i^{'},j^{'})},
\label{e1}
\end{equation}
where the denominator is the distance between the sites $(i,j)$ and
$(i^{'},j^{'})$, the sum is restricted to $(i,j) \neq (i^{'},j^{'})$,
and 
\begin{equation}
\delta(i,j,i^{'},j^{'})=
\left\{
\begin{array}{ll}
x|i-i^{'}|,\;\;\mbox{if $j=j^{'}$},\\ \mbox{} \\
%d_{i,i^{'}}=
\sqrt{x^2+f_i^2+f_{i^{'}}^2-2f_if_{i^{'}}\cos\theta},
\;\;\mbox{if $j \neq j^{'}$;}
\end{array}
\right.
\label{e2}
\end{equation}
with $x=d/b$ and
\begin{equation}
f_i=\frac{Z+1-2i}{2}.
\label{e3}
\end{equation}
The values of occupation variables obey the constraints $\sum_{i=1}^Z
\sigma_{i1}=\sum_{i=1}^Z\sigma_{i2}=n$.

The partition function of the model is given by
\begin{equation}
Q={\sum_{\{\sigma\}}}' \exp(-\beta {\cal H})=
{\sum_{\{\sigma\}}}' \exp(-\xi H),
\label{e4}
\end{equation}
where the prime denotes the constraint of fixed numbers of condensed
counterions on each polyion. The adimensional reduced Hamiltonian $H$
is given by
\begin{equation}
H=\frac{1}{2}\sum_{(i,j) \neq (i^{'},j^{'})}
\frac{(1-\alpha \sigma_{ij})(1-\alpha \sigma_{i^{'}j^{'}})} 
{\delta(i,j,i^{'},j^{'})},
\label{e5}
\end{equation}
and $\xi=\beta q^2/Db$ is the Manning parameter \cite{man69}.
For a given counterion configuration $\{\sigma\}$, if a
transformation $\sigma^{'}_{ij}=1-\sigma_{ij}$ is performed, the
Hamiltonian changes as
\begin{equation}
H(Z,\alpha,n,x,\theta,\{\sigma\})=(\alpha-1)^2
H(Z,\alpha^{'},n^{'},x,\theta,\{\sigma^{'}\}),
\label{e6}
\end{equation}
where $\alpha^{'}=\alpha/(\alpha-1)$ and $n^{'}=Z-n$.
This relation leads to the following invariance property of the
partition function of the model 
\begin{equation}
Q(Z,n,\xi,\alpha,x,\theta)=
Q(Z,Z-n,[\alpha-1]^2\xi,\alpha/[\alpha-1],x,\theta),
\label{e7}
\end{equation}
and therefore we may restrict ourselves to $n<Z/2$ in the
calculations. The free energy of the model is $\phi=-k_B T \ln Q$, 
and
the force between the rods is $F=-\frac{\partial \phi}{\partial
d}$. It is then useful to define an adimensional force as
\begin{equation}
f=\frac{Db^2F}{q^2}=\frac{1}{\xi Q} \frac{\partial Q}{\partial x}.
\label{e8}
\end{equation}
It should be stressed that the definition of $f$ we use here is
different from the one used in \cite{are99}, since the earlier
definition diverges at vanishing temperature. The sign was chosen in
such a way that repulsive forces are positive.

To solve the model exactly, one may define activity variables
\begin{eqnarray}
y_i & = & \exp\left[-\frac{\xi}{i}\right], i=1,\ldots,Z-1\\
z_{ij} & = & \exp\left[-\frac{\xi}{d_{ij}}\right], i,j=1,\ldots,Z,
\label{e9}
\end{eqnarray}
where $d_{ij}=\sqrt{x^2+f_i^2+f_{i'}^2-2f_if_{i'}\cos\theta}$.
It may then be noticed that the partition function may be written
as
\begin{equation}
Q=\sum_{i=1}^{N_c} w_i, 
\label{e10}
\end{equation}
where
\[
N_c=\left[\frac{Z!}{n!(Z-n)!}\right]^2
\]
is the number of condensed counterions configurations and the
statistical weight of the $i$'th configuration is given by
\begin{equation}
w_i=\prod_{j=1}^{Z-1} y_j^{u_{ij}} \prod_{k,l=1}^Z
z_{kj}^{v_{ikl}},
\label{e11}
\end{equation}
$u_{ij}$ and $v_{ikl}$ being quadratic polynomials in $\alpha$
with integer coefficients. It is possible to generate these sets of
integer numbers with a computer program and thus obtain the partition
function of the model exactly. On a conventional personal computer
with a rather moderate processing time it is easy to obtain results up
to $Z=13$, and simulations \cite{are99} for larger polyions show that
the qualitative behavior of the model does not change much beyond
this value, so we will restrict ourselves here to $Z \leq 13$. 

\section{Results for the thermodynamic behavior of the model}
\label{III}
\begin{figure}
\centerline{\epsfig{file=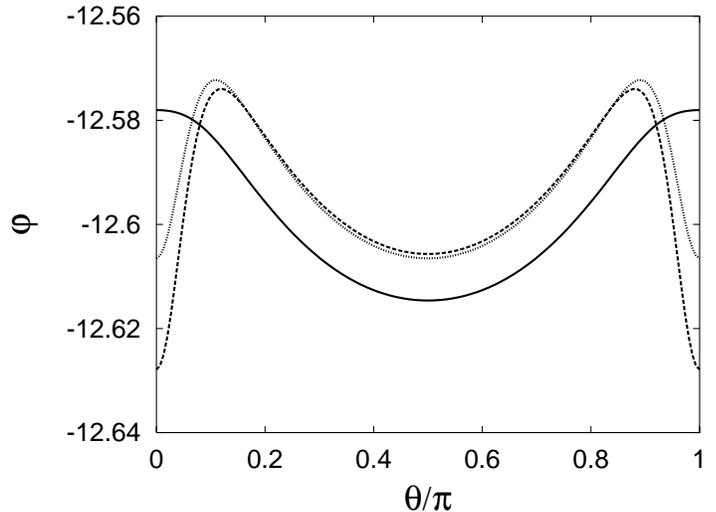,width=7cm,angle=270} }
\caption{Free energy $\varphi$ as a function of the angle $\theta$ for
$Z=9$, $n=4$, $a=0.8$, $\xi=2$. The curves shown are for: $x=2$ (full
line), $x=1.3676010335$ (dotted line), and $x=1.3$ (dashed line).
Notice that the free energies for $\theta=0$ and $\theta=\pi/2$ are
equal in the second case.}
\label{f2}
\end{figure}

The thermodynamic behavior of the model is determined by the free
energy $\phi$, and we will start considering the
dependence of $\phi$ on  $\theta$. For convenience, we
define an adimensional free energy
\begin{equation}
\varphi=\frac{Db}{q^2}\phi=-\frac{1}{\xi}\ln Q.
\label{e12}
\end{equation}
The free energy is a function of the parameters $Z$, $n$, $\xi$,
$\alpha$, $x$, and $\theta$. For reasons which will become clear
below, we replace the parameter $\alpha$ by
\begin{equation}
a=\frac{2n\alpha}{Z}-1,
\label{e13}
\end{equation}
so that we will consider the free energy
$\varphi(Z,n,\xi,a,x,\theta)$.  For all cases we
noticed that the global minimum of the free energy is located either
at the parallel ($\theta=0$) or at the perpendicular ($\theta=\pi/2$)
configuration of the rods. An example of this is shown in 
 Fig.~\ref{f2}, and one may note that, in general, the parallel
configuration is stable for small distance $x$ and the perpendicular
one becomes stable as $x$ is increased. So, in what follows we will
concentrate our attention on the parallel and perpendicular rod
configurations only.

We will now focus on the regions of the parameter space
where the parallel and perpendicular configurations are stable. We
will start discussing the behavior of the model at vanishing values
of $x$. In this limit, the partition sum is dominated by the
contributions coming from interactions between charges separated by a
distance $x$.  For the case $\theta=0$ we have $Z$ pairs in this
situation. We may then rewrite the partition
function as
\begin{equation}
Q_{\parallel} = \sum_{i=1}^{N_c}\prod_{j=1}^{Z-1}y_j^{u_{ij}}
z_0^{\sum_{k=1}^{Z}v_{ikk}} 
\prod_{l \neq m} z_{lm}^{v_{ilm}},
\label{e14}
\end{equation}
where $z_0=z_{kk}=\exp(-\xi/x)$ vanishes as $x \to 0$. Now let us
call 
\begin{equation}
v=\min_{i} v_i=\min_{i}\sum_{k=1}^{Z} v_{ikk},
\label{e15}
\end{equation}
and let us suppose that the first $N_1$ of the $N_c$ condensed
counterion configurations correspond to this value of $v_i$. Thus
\begin{equation}
Q_{\parallel}=z_0^v W_{\parallel}(1+P),
\label{e16}
\end{equation}
where 
\[
W_{\parallel}=\sum_{i=1}^{N_1} \prod_{j=1}^{Z-1} y_j^{u_{ij}}
\prod_{l \neq m} z_{lm}^{v_{ilm}},
\]
and
\[
P=\frac{1}{W_1} \sum_{i=N_1+1}^{N_c} z_0^{v_i-v} \prod_{j=1}^{Z-1}
y_j^{u_{ij}}  \prod_{l \neq m} z_{lm}^{v_{ilm}}
\]
vanishes as $x \to 0$, since $v_i>v$. The $N_1$ configurations
considered here are the ones that maximize the number of $(-1,-1+\alpha)$ 
pairs
separated by a distance $x$, and thus $v=Z-2n\alpha=-Za$. For the
parallel configuration, we thus have that at small values of $x$, the
partition function is asymptotically given by
\begin{equation}
Q_{\parallel} \approx W_{\parallel} z_0^{-Za}.
\label{e17}
\end{equation}
A simple combinatorial calculation leads to
\begin{equation}
N_1=\frac{Z!}{(n!)^2(Z-2n)!}.
\label{e17p}
\end{equation}

The same line of reasoning may be applied to the perpendicular case,
where only the pair of central charges $([Z+1]/2,1)$ and $([Z+1]/2,2)$ is
separated by a distance $x$. One thus has in this case
\begin{equation}
Q_{\perp} \approx W_{\perp} z_0^{1-\alpha},
\label{e18}
\end{equation}
with
\begin{equation}
W_{\perp}=\sum_{i=1}^{N_2} \prod_{j=1}^{Z-1} y_j^{u_{ij}}
\prod_{(l,m)
\neq ([Z+1]/2,[Z+1]/2)} z_{lm}^{v_{ilm}},
\label{e19}
\end{equation}
where the first $N_2$ configurations are now supposed to be the ones
with a central $(-1,-1+\alpha)$ pair, and
\begin{equation}
N_2=\frac{[(Z-1)!]^2}{(Z-n)!(Z-n-1)!n!(n-1)!}.
\label{e19p}
\end{equation}
>From equations~\ref{e17} and~\ref{e18}, $Q_{\parallel}= Q_{\perp}$ at
small distances leads to
\begin{equation}
a=a_0+a_1x+{\cal O}(x^2),
\label{e20}
\end{equation}
with
\begin{eqnarray}
a_0 & = & \frac{Z-2n}{Z(2n-1)},\\
a_1 & = & \frac{2n}{Z\xi(2n-1)}(\ln W_2-\ln W_1).
\label{e21}
\end{eqnarray}
It is thus apparent that the sign of the inclination of the curve
$\varphi_{\parallel}=\varphi_{\perp}$ in the $(a,x)$ plane 
at $x=0$ is determined by the
sign of $W_2-W_1$, since $\ln(x)$ is a 
monotonically increasing function.

\begin{figure}
\centerline{\epsfig{file=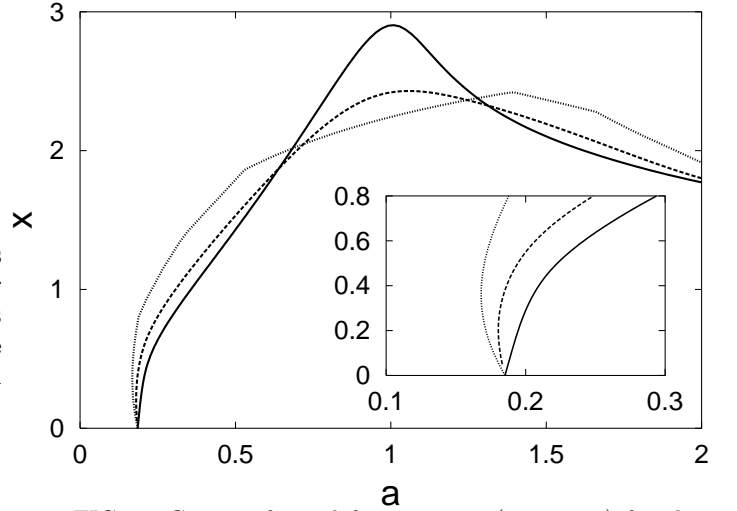,width=7cm,angle=270} }
\caption{Curves of equal free energies
($\varphi_{\parallel}=\varphi_{\perp}$) for the model with $Z=9$,
$n=2$, for $\xi=1$ (full line), $\xi=2$ (dashed line) and $\xi \to
\infty$ (dotted line). The parallel configuration is stable below 
the curves. Notice, in the inset,  the sign of $a_1$ varies with $\xi$.
in these cases.}
\label{f3}
\end{figure}

\begin{figure}
\centerline{\epsfig{file=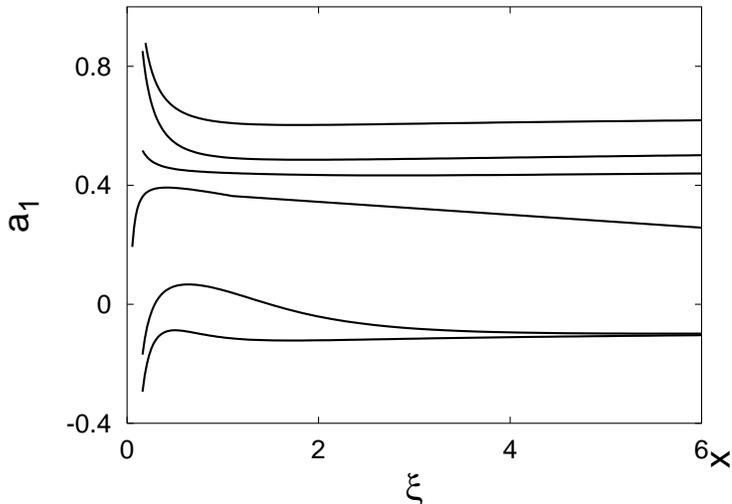,width=7cm,angle=270} }
\caption{Inclination of the curves of equal free energies, $da/dx=a_1$,
at $x=0$,
 as a function of $\xi$ for some cases $(Z,n)$ from top to
bottom: $(13,4)$, $(9,3)$, $(11,3)$, $(13,3)$, $(9,2)$, $(11,2)$.}
\label{f4}
\end{figure}

In Fig.~\ref{f3} the curve of equal free energies in the plane
$(a,x)$ is shown for the case $Z=9$, $n=2$. For all temperatures, the
curves meet at $(a_0=5/27,x=0)$, as expected. For the case of the
ground state (vanishing temperature), the curve displays
discontinuous derivatives at points where the configurations of the
ground state change. Another point which is worth observing is that
for $\xi>1.489528...$ or $\xi<0.288876...$ the curve starts
with {\em negative} inclination at $(a_0,0)$. Thus, for a value of
$a$ somewhat smaller than $a_0$ the perpendicular configuration is
stable at large {\em and} small distances, while parallel rods are
stable at intermediate distances. Thus, in this case we find that the
perpendicular phase is reentrant as the distance $x$ is lowered.
Figure~\ref{f4} shows the initial inclination $a_1=\left(\frac{\partial
a}{\partial x}\right)_{x=0}$ as a function of $\xi$ for some
examples. As is apparent in expression~\ref{e21}, these inclinations
diverge as $\xi \to 0$, their sign in this limit being determined by
$W_2-W_1$. As a general rule, one notices that reentrant behavior is
found for relatively small values of $n$ in the examples that 
we have studied ($n<3$ for $Z=7,9,11$ and $n<4$ for $Z=13$).

\begin{figure}
\centerline{\epsfig{file=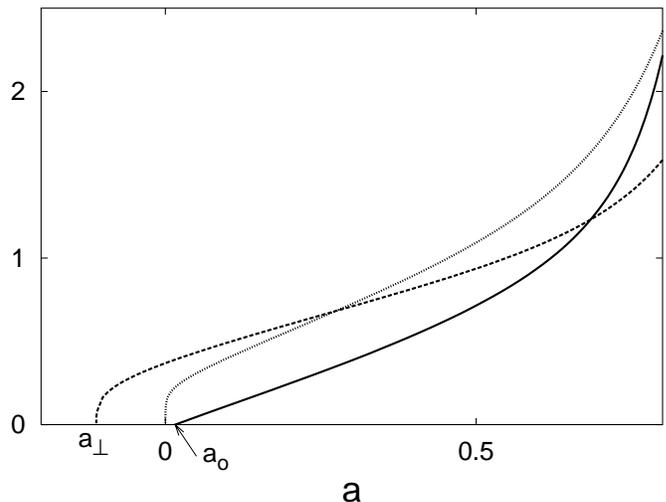,width=7cm,angle=270} }
\caption{Curves of equal free energies (full line), $f_{\parallel}=0$
(dotted line), and $f_{\perp}=0$ (dashed line), for the model with
$Z=9$, $n=4$, and $\xi=1$. Note that bellow the solid curve
the parallel configuration has the lowest free energy, while
above the  perpendicular configuration is energetically favored.
The parallel configuration has an attractive force between the polyions
bellow the dotted line.  Similarly, the perpendicular configuration 
is attractive
bellow the dashed line.}
\label{f5}
\end{figure}

Finally, we discuss the behavior of the force defined in 
Equation~\ref{e8}. As a general rule, for sufficiently large values of $a$,
the force is repulsive at large distances and becomes attractive as
the distance is lowered. It is useful to find the value of the
distance $x$ for which the forces $f_{\parallel}$ and $f_{\perp}$
vanish. The behavior of these curves at small distances may be found
from the asymptotic behavior of the free energies described in
equations \ref{e17} and \ref{e18}. We find 
\begin{equation}
f_{\parallel} \approx h_{\parallel}x-\frac{Za}{x^2},
\label{e22}
\end{equation}
where
\[ 
\frac{1}{\xi W_{\parallel}} \frac{\partial W_{\parallel}}{\partial x}
=h_{\parallel}x+{\cal O}(x^2).
\]
Thus, for small values of $x$, the curve $f_{\parallel}=0$ reads
\begin{equation}
x \approx \left( \frac{Za}{h_{\parallel}} \right)^{1/3},
\label{e23}
\end{equation}
so that attractive forces in the parallel configuration are possible
only if $a>0$, since $h_{\parallel}>0$. For the perpendicular configuration, similar
considerations lead to 
\begin{equation}
f_{\perp} \approx h_{\perp}x-\frac{\alpha-1}{x^2},
\label{e24}
\end{equation}
and therefore the curve $f_{\perp}=0$ for $x \ll 1$ is given by
\begin{equation}
x \approx \left( \frac{\alpha-1}{h_{\perp}} \right)^{1/3}.
\label{e25}
\end{equation}
For small separations between the rods, the curve $f_{\perp}=0$,
in the $(a,x)$ plane, tends to
$a=a_{\perp}=-1+2n/Z$.  In Fig.~\ref{f5} the curves
$f_{\parallel}=0$, $f_{\perp}=0$, and
$\varphi_{\parallel}=\varphi_{\perp}$ are depicted for a particular
case. One notices that for $a>a_0$, the force between the rods is
attractive for sufficiently small distances $x$ and the rods are in
the parallel configuration. For $a_{\perp}<a<a_0$ the force is still
attractive at small $x$, but the rods are in the perpendicular
configuration.

\begin{figure}
\centerline{\epsfig{file=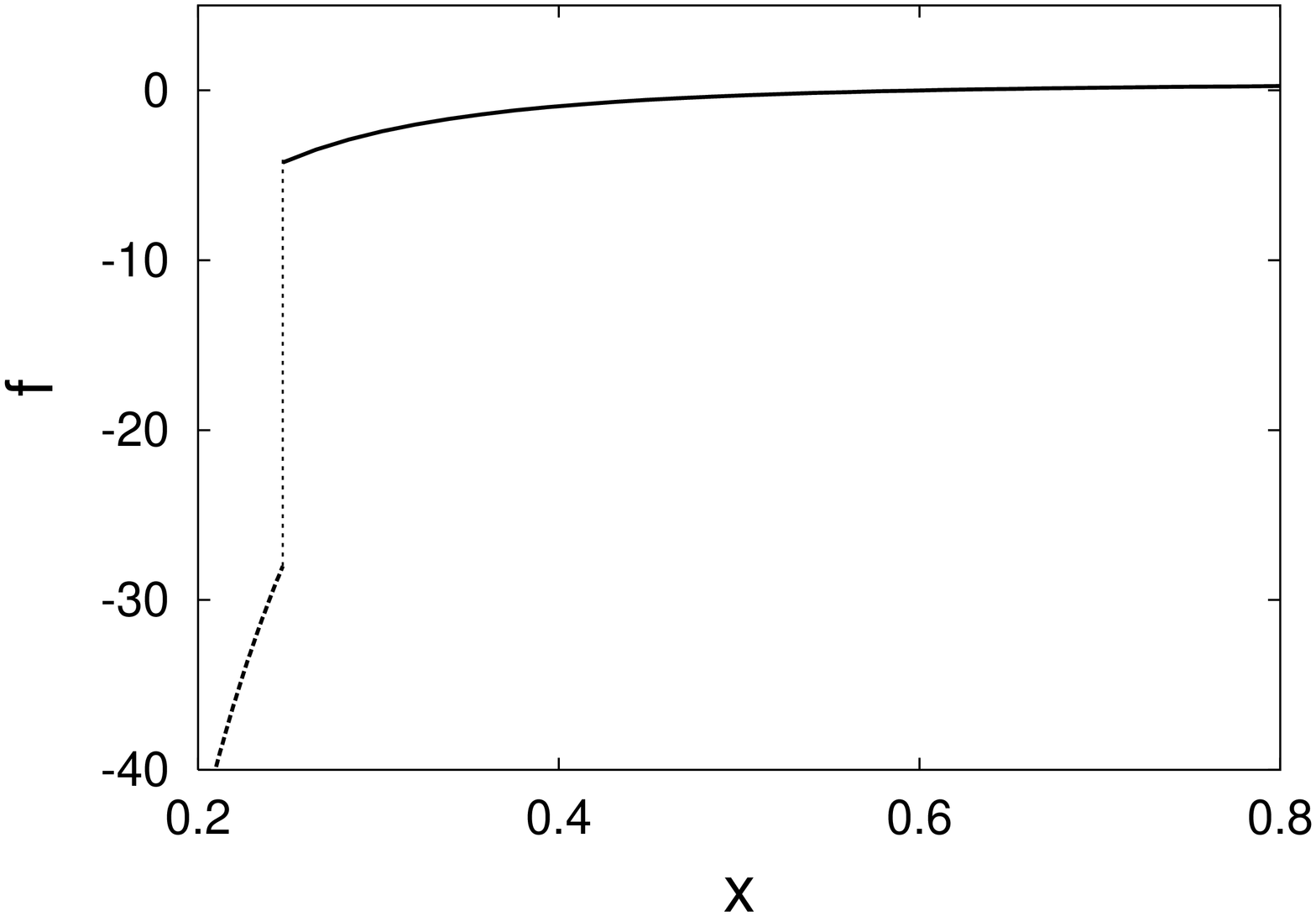,width=7cm,angle=270} }
\caption{Force as a function of the distance between rods for $Z=9$,
$n=4$, $a=0.2$ and $\xi=1$. The full line corresponds to
perpendicular rods and the dashed line to parallel rods.}
\label{f6}
\end{figure}

\begin{figure}
\centerline{\epsfig{file=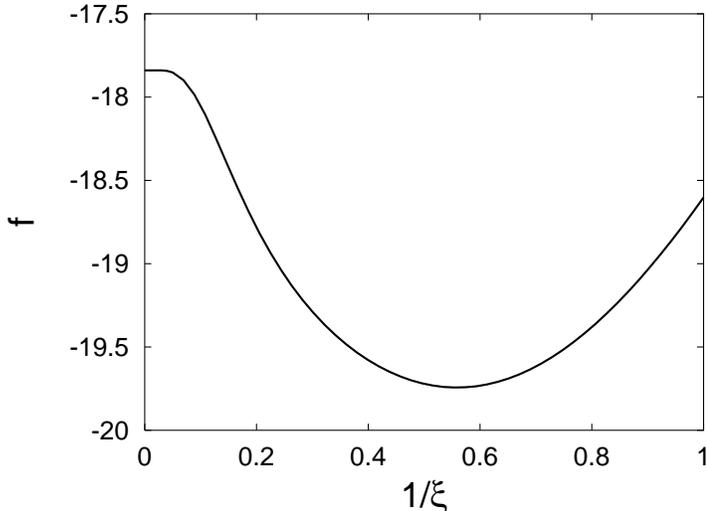,width=7cm,angle=270} }
\caption{Force as a function of the temperature $1/\xi$ for $Z=9$,
$n=4$, $a=0.6$, and $x=0.5$.}
\label{f7}
\end{figure}

The behavior of the force as a function of the distance $x$ is shown
in Fig.~\ref{f6} for $a=0.2$. At $x=1$ the force is
repulsive and the rods are perpendicular. As the distance is lowered,
the force becomes attractive, with the rods still in the
perpendicular configuration. At $x=0.246924...$ the rods change to the
parallel configuration and the attractive force becomes much larger.
At a small distance, the force displays an asymptotic behavior of the form
$-Za/x^2$, according to Eq.~\ref{e22}. At $a=0.75$ the force is
repulsive at large distances, with perpendicular rods. As the distance
is lowered, the force changes discontinuously to attractive as the
rods become parallel. Finally, an example of the behavior of the
force as a function of the temperature $1/\xi$ may be seen in 
Fig.~\ref{f7}. For the values of the parameters used to obtain these data,
the parallel configuration is stable. In general, the modulus of the
force increases as the temperature is lowered, since the charge
correlations grow in this case. However, at relatively low
temperatures and short separations between the polyions, 
this rule may not apply, as is the case in the
example shown. The reason for this is that the ground state
configuration, which corresponds to the lowest electrostatic energy, 
is not, in
general, the configuration which maximizes the attractive force.
For low temperature and short distances  
there are configurations which have forces more attractive than what
is found in the ground state.  The total force, being a weighted
mean of forces associated with all configurations can, therefore, become
more attractive than the force at T=0.

\section{Discussion and conclusions}
\label{IV}

To get a better understanding of the range of 
electrostatic correlational forces involved, we study
a simple example of two parallel long polyions with $Z/2$
condensed divalent counterions.  In the ground state
the counterions are distributed periodically along
the polyions. At finite temperature, this
periodicity will be destroyed, the correlations
between the condensed counterions, however, will 
persist. Thus, for sufficiently large electrostatic 
coupling, i.e. Manning parameter, 
the electrostatic  potential
at position ($\rho,s$) from a polyion can be approximated
as,
\begin{equation}
\phi(\rho,s) = q\sum_{n=-\infty}^{\infty} \frac{(-1)^n}{\sqrt{\rho^2
+(s-nb)^2}} \;.
\end{equation}
Appealing to Poisson sum rule the asymptotic large distance behavior
or this sum can be evaluated yielding 
\begin{equation}
\phi(\rho,s) = \frac{4q}{\sqrt{2\rho b}} \exp\left(-\frac{\pi\rho}{b}
\right) \cos \left(\frac{\pi s}{b}\right) \;.
\end{equation}
The energy of interaction between two lines of charge with 
staggered charges is then
\begin{equation}
E=- \frac{4qZ}{\sqrt{2\rho b}} \exp\left(-\frac{\pi\rho}{b}
\right)
\end{equation}
and the force, $F=-\partial E/\partial \rho$, is
\begin{equation}
\frac{F}{Z} = - \frac{4q}{\sqrt{2\rho b^3}}  \exp\left(-\frac{\pi\rho}{b}
\right) \left( \pi + \frac{b}{2\rho}\right)
\end{equation}
We see that the correlation induced attraction decays exponentially,
with the characteristic range of $l=b/\pi$.

We have studied the electrostatic interaction between two charged rods
with a layer of condensed counterions. This is the simplest model of
interaction for like-charged polyions in a polyelectrolyte solutions.
It is shown that in spite of the equal net charge on the two macromolecules
the correlations between the condensed counterions can produce an 
effective attraction.
We show, however, that this attraction is of extremely short range,
comparable to the monomer separation along the macromolecules.   
Furthermore, it is found that at large distances the monopolar repulsion
between the charge densities forces the polyions into a perpendicular
configuration.  At short distances the correlations between the condensed
counterions can become sufficient to produce a macromolecular alignment.
The energy barrier associated with the transition from the perpendicular
to the parallel configuration might be relevant for the kinematics of bundle
formation in solutions of stiff polyelectrolytes~\cite{ha99}.

%\end{thebibliography}

\end{multicols}
\end{document}